*Article*

# The Trail Making Test in Virtual Reality (TMT-VR): The Effects of Interaction Modes and Gaming Skills on Cognitive Performance of Young Adults

**Evgenia Giatzoglou** [1,†]**, Panagiotis Vorias** [1,†]**, Ryan Kemm** [1]**, Irene Karayianni** [1]**, Chrysanthi Nega** [1]
**and Panagiotis Kourtesis** [1,2,3,4,*]

[1] Department of Psychology, The American College of Greece, 15342 Athens, Greece;
  e.giatzoglou@acg.edu (E.G.); p.vorgias@acg.edu (P.V.); r.kemm@acg.edu (R.K.); ikarayianni@acg.edu (I.K.); cnega@acg.edu (C.N.).
[2] Department of Informatics & Telecommunications, National and Kapodistrian University of Athens,
  16122 Athens, Greece
[3] Department of Psychology, National and Kapodistrian University of Athens, 15784 Athens, Greece
[4] Department of Psychology, The University of Edinburgh, Edinburgh EH8 9Y, UK
[*] Correspondence: pkourtesis@acg.edu
[†] These authors contributed equally to this work.

**Featured Application: The Trail Making Test in Virtual Reality (TMT-VR) provides a highly adaptable and immersive tool for cognitive and neuropsychological assessments. Its application extends beyond traditional clinical settings, offering potential use in remote assessments, telemedicine, and longitudinal studies that require repeated measures. The fully automated, randomized, and naturalistic environment of the TMT-VR ensures precise and consistent evaluation across diverse populations, making it particularly useful for assessing cognitive functions in older adults, individuals with neurological conditions, and populations with limited technological exposure. Additionally, the TMT-VR could serve as a valuable tool in cognitive rehabilitation programs, where immersive and engaging tasks are crucial for patient motivation and adherence. However, validation studies are required for scrutinizing its utility in older adults, clinical populations, and settings.**

**Abstract:** Virtual Reality (VR) is increasingly used in neuropsychological assessments due to its ability to simulate real-world environments. This study aimed to develop and evaluate the Trail Making Test in VR (TMT-VR) and investigate the effects of different interaction modes and gaming skills on cognitive performance. A total of 71 young female and male adults (aged 18–35) with high and low gaming skills participated in this study. Participants completed the TMT-VR using three interaction modes as follows: eye-tracking, head movement, and controller. Performance metrics included task completion time and accuracy. User experience, usability, and acceptability of TMT-VR were also examined. Results showed that both eye tracking and head movement modes significantly outperformed the controller in terms of task completion time and accuracy. No significant differences were found between eye tracking and head movement modes. Gaming skills did not significantly influence task performance using any interaction mode. The TMT-VR demonstrates high usability, acceptability, and user experience among participants. The findings suggest that VR-based assessments can effectively measure cognitive performance without being influenced by prior gaming skills, indicating potential applicability for diverse populations.

## 1. Introduction

Virtual Reality (VR) has emerged as a transformative tool in neuropsychological assessment, enabling the creation of immersive, three-dimensional environments that simulate real-world scenarios with high fidelity. Initially developed to enhance gaming experiences, VR has, over the past two decades, been increasingly adopted in research, particularly for replicating environment-specific stimuli within controlled laboratory contexts that were previously difficult to access [1–4]. This growing interest in VR-based methodologies has been particularly notable in the assessment and enhancement of cognitive [5] and motor functions [6] through both immersive and non-immersive technologies. Its utility in psychological research is underscored by its naturalistic interaction capabilities, which allow for the presentation of complex, real-world tasks in a controlled, repeatable manner. This has proven invaluable for applications in rehabilitation, cognitive training, and neuropsychological assessment [7,8].

Extensive research, including our own, has demonstrated the efficacy of VR in neuropsychological assessments across various populations, such as children with attention deficit hyperactivity disorder (ADHD) and adults with autism spectrum disorder (ASD). These studies suggest that VR provides a flexible, adaptable environment that can be customized to minimize or incorporate external distractions, creating a safe, engaging setting conducive to assessment [9,10]. Additionally, VR-based tools offer a platform not only for diverse behavioral and cognitive tests but also for enhancing and measuring critical cognitive functions such as working memory, executive functioning, and behavior [11–13]. Moreover, these tools are positioned to be both cost-effective and time-efficient, potentially revolutionizing the landscape of cognitive assessment, diagnosis, and rehabilitation by making these processes more accessible and scalable.

### 1.1. Limitations of Traditional Cognitive Assessment Tools

Traditional neuropsychological assessments, such as the Stroop Test and Trail Making Test (TMT), are typically administered through paper-and-pencil or computer-based formats. The Trail Making Test (TMT), in particular, is a widely used neuropsychological tool that assesses cognitive functions like visual attention, task switching, and cognitive flexibility. It consists of two parts as follows: TMT-A, which requires connecting a series of numbered circles in ascending order, measuring sequencing and processing speed, and TMT-B, which requires alternating between numbers and letters, providing a measure of more complex cognitive processes such as mental flexibility and executive functioning [14]. However, these methods often fail to capture the complexity of real-world environments, lacking the distractions and stressors that individuals encounter in daily life. This limitation diminishes the ecological validity of these assessments, meaning they may not fully reflect an individual's cognitive abilities in everyday settings [15]. In contrast, VR tools excel at integrating dynamic perceptual stimuli with realistic scenarios within controlled laboratory settings, thereby significantly enhancing ecological validity [1,2].

Another significant drawback of traditional assessments is their reliance on the presence of a healthcare professional and their typical use only after cognitive impairments have been identified, potentially hindering early intervention [15]. Moreover, these kinds of assessments often cannot provide real-time feedback or continuous monitoring, making them ineffective in assessing cognitive changes over a period of time or in comparing the effectiveness of current interventions [16]. VR-based tools, however, can simulate real-world conditions while retaining the controlled aspects of traditional methods. This capability not only facilitates earlier and potentially remote or self-administered evaluations but also broadens access to neuropsychological assessments, thus promoting early intervention and more personalized care [2,17].

One of VR's most notable advantages in neuropsychological assessments is its ability to reduce human error in both test administration and result calculation. Traditional assessments, often perceived as monotonous and repetitive, can suffer from data quality issues due to participant disengagement. VR offers a more immersive and engaging alternative, which can enhance user participation and data accuracy [10,15]. Furthermore, compared to traditional neuropsychological tools, VR assessments can better discriminate between different levels of cognitive impairment (such as mild cognitive impairment) due to improved sensitivity and specificity [18]. Moreover, VR tools allow for the collection of diverse data types during assessments, including eye and physical movements and facial expressions [11,15,19]. This multidimensional data collection enables a more holistic evaluation, potentially revealing cognitive aspects that traditional methods may overlook. Additionally, VR technology exhibits potential for real-time adaptation to the user's performance, providing a dynamic assessment environment [20], providing millisecond-level precision in performance recording, and enabling the calculation of detailed performance metrics, such as means and standard deviations, ensuring high precision and reliability in cognitive assessments [11,21].

## 1.2. Executive Functions and the Role of Gaming Skills

Executive functions encompass fundamental processes such as mental set shifting, information updating, and monitoring, as well as inhibitory control [22], which form the basis for higher-order cognitive abilities like reasoning, problem-solving, decision-making, and planning. These functions are critical for goal-directed behavior and adaptability in complex, dynamic environments. In recent years, video gaming has emerged as a valuable tool in cognitive science, offering insights into the mechanisms underlying executive and cognitive functions. Action video games (AVGs), including first-person (FPS) and third-person shooter games, have been shown to enhance attentional control, visuospatial processing, and cognitive flexibility—skills that are essential for both everyday functioning and performance on neuropsychological assessments [23].

Empirical research has consistently demonstrated that AVG players exhibit superior mental rotation abilities, faster task switching, and enhanced working memory compared to non-gamers [24,25]. For example, AVG gamers outperform non-gamers on the traditional Trail Making Test (TMT), particularly in terms of task completion times, indicating heightened attentional control, psychomotor speed, and error inhibition [15,26]. Moreover, studies comparing gamers who engage in different types of video games have found that those who play action games achieve faster visual scanning scores on TMT-A compared to puzzle game players, though no significant differences are observed on TMT-B, suggesting that the influence of gaming on mental flexibility may be limited [27].

Despite these positive associations, the literature presents a nuanced view of the impact of video gaming on cognitive performance. While some studies suggest that video games enhance visual and selective attention and promote efficient spatial learning strategies, others indicate that gaming may lead to fragmented attention and increased decision-making errors, particularly in tasks requiring suppression of unwanted responses [28]. This dichotomy suggests that while gaming may enhance certain aspects of executive function, such as working memory and attentional load, it may simultaneously impair others, such as inhibitory control.

Furthermore, there is evidence to suggest that video gamers may behave differently in virtual reality (VR) environments compared to non-gamers, likely due to their lower emotional involvement and greater familiarity with immersive technologies [29]. This behavioral difference underscores the importance of considering individual differences in technology proficiency and gaming experience when assessing cognitive performance in VR.

It is important to note that while gaming skills may influence performance on VR-based assessments, this effect has predominantly been observed in controller-based interactions. However, when employing more naturalistic and ergonomic interaction methods, such as eye tracking or head movement, the impact of gaming proficiency on performance appears to be diminished, allowing gamers and non-gamers to perform on par with one another [11]. This finding is particularly relevant for the development of VR-based cognitive assessments, as it suggests that these tools can be designed to be accessible and effective for a broad population, regardless of prior gaming experience.

## 1.3. Importance of Ecological Validity in Neuropsychological Assessments

Virtual Reality (VR) offers unprecedented potential in neuropsychological research by enhancing the ecological validity of assessments. Ecological validity refers to the extent to which a neuropsychological tool replicates real-world conditions (verisimilitude) and the degree of correlation between the observed performance in this study and real-life behavior (veridicality), thereby facilitating the generalization of results to everyday activities [2,30]. The ecological validity of VR tools is anchored in their ability to seamlessly integrate dynamic perceptual stimuli with the controlled conditions of laboratory settings, enhancing the realism of the user experience [1]. By allowing researchers to create customized environments tailored to specific users and contexts, VR enables the replication of complex everyday stimuli and the real-time study of behavioral responses, leading to more reliable outcomes [2,31].

## 1.4. Cybersickness

Cybersickness, which manifests as nausea, dizziness, and disorientation, poses a significant challenge in the deployment of Virtual Reality (VR) technologies. This condition is akin to motion sickness and is often triggered by sensory conflicts inherent in immersive VR environments. Factors such as the ergonomic design of VR headsets—particularly their weight and fit—and software-related issues like improper calibration can exacerbate these symptoms [32]. The presence of cybersickness not only impacts the user's comfort but can also compromise the reliability of cognitive assessments conducted in VR, making it a critical issue to address [1,2,32–34].

Recent advancements in VR technology have made significant strides in reducing the incidence and severity of cybersickness. Innovations in hardware, including better weight distribution and enhanced visual fidelity, along with software improvements like teleportation-based movement systems and optimized user interfaces, have contributed to a more comfortable and immersive experience [11,35,36]. These developments are essential for ensuring that VR tools are not only effective but also accessible to a broader population. By minimizing cybersickness, these improvements enhance the reliability and applicability of VR-based cognitive assessments, making them more enjoyable and feasible for users across different demographics [2,35–37].

*1.5. The Trail Making Test in Virtual Reality*

The Trail Making Test (TMT) is a widely recognized neuropsychological assessment tool used to measure executive functioning and cognitive performance. It is particularly effective in distinguishing between healthy aging processes and pathological conditions and is commonly employed in the early screening for Alzheimers disease [38]. The TMT comprises two parts as follows: TMT-A, which assesses attention, visual scanning, and motor speed, and TMT-B, which evaluates cognitive flexibility, working memory, set-shifting, and inhibitory control [14,39].

Developing the Trail Making Test in Virtual Reality (TMT-VR) introduces a novel dimension to cognitive assessment. This adaptation seeks to maintain the traditional cognitive measurements of the TMT while exploring how these abilities are affected when targets are spatially distributed within a 3D environment. The immersive nature of VR offers significant insights into spatial cognition, attention, and the impact of interaction modalities on cognitive performance [10,40,41]. However, a thorough investigation is required to ensure that the TMT-VR will be an appropriate and effective neuropsychological tool that can be implemented in diverse populations.

### 1.5.1. Interaction Modes

The ability for users to frequently shift their body position is a defining feature of 3D virtual environments, distinguishing them from traditional 2D scenes. Various interaction methods are available for engaging with VR environments, ranging from sensors, keyboards, and voice input in less immersive scenarios to controllers, hand motions, eye-tracking, and head movement in fully immersive VR (IVR) [35,42]. Given the specific requirements of the Trail Making Test in Virtual Reality (TMT-VR), which necessitates selecting targets from a distance within an immersive environment, the choice of appropriate interaction methods was critical. The TMT-VR, initially developed for assessing cognitive functions in older adults, necessitates interaction modes that minimize the need for extensive body movement or prior technological experience. Hands-free interaction methods, such as head movement and eye-tracking, are particularly promising for this demographic, as they require minimal muscle effort and offer accessible input options [41,43,44].

Eye-tracking, in particular, can be invaluable in scenarios where other modalities are unavailable or impractical, such as when users' hands are occupied or when severe motor disabilities are present [45]. Recent studies have highlighted that physiological signals like eye movements are both intuitive and natural, with a strong association between eye tracking and immersion in VR [46]. Furthermore, research suggests that eyes naturally rotate faster than the head, making eye-gaze interaction quicker when it comes to task completion time compared with head–gaze interaction, especially in tasks with a larger field of view [47]. Lastly, eye tracking has also demonstrated higher accuracy in tasks involving depth perception [48].

Conversely, head–gaze interaction is noted for its superior precision and stability, making it more effective for tasks requiring exact manipulation. Studies have shown that head-only interaction results in the fastest selection times and accuracy compared to eye-only and hybrid modalities, with participants expressing a preference for its comfort and precise response [49]. Despite their distinct strengths, eye tracking and head–gaze interaction modes are considered complementary or interchangeable depending on the context. However, it has been suggested that significant technological advancements are needed for eye tracking to fully replace head selection due to inherent physiological characteristics and current limitations in eye-tracking data quality [50–52]. While eye tracking and head–gaze interaction offer minimal effort and intuitive use, the controller input is a widely used method in VR environments due to its familiarity and tactile feedback, making it essential to compare its effectiveness against hands-free methods. Understanding the impact of these different interaction modes can help optimize TMT-VR for diverse user populations, including those accustomed to traditional VR controls.

### 1.5.2. Acceptability, Usability, and User Experience

To consider TMT-VR as an innovative assessment tool necessitates a thorough evaluation of its usability, acceptability, and overall user experience. Following User Centered Design (UCD) principles, it is essential to ensure that the tool meets the needs and expectations of its intended users, ultimately leading to more accurate, enjoyable, and reliable cognitive assessments [10,35,41].

In VR, usability refers to the efficiency and ease with which participants can engage with the system to achieve their goals. This encompasses factors such as the perceived user-friendliness of the equipment and the ability to prevent errors during tasks, which is closely related to the capacity to ignore distractions and inhibit automatic responses [10,53].

Acceptability is defined as the degree to which participants are willing to use VR technology and equipment. This includes considerations of comfort and social acceptability when engaging with VR software [54]. User experience (UX) refers to the emotional aspects of usability, particularly the sense of satisfaction, achievement, and engagement. It encapsulates the sensory, cognitive, emotional, and behavioral responses evoked by the virtual experience, including immersion, presence, realism, and enjoyment [10,53].

When designing a new VR tool, these factors must be quantitatively assessed to ensure a high-quality user experience [55]. Recent VR systems developed for various experimental contexts have achieved high scores in acceptability, usability, and perceived effectiveness across diverse populations, including cancer patients, older adults, hospital staff, and primary caretakers [12,56]. Furthermore, VR has been proven effective in managing pain under experimental conditions, highlighting the potential of tools like the TMT-VR to deliver significant benefits in neuropsychological assessments [57].

### 1.6. Objectives and Hypotheses of the Current Study

Building on the considerations outlined above, the development of the TMT-VR aims to be the first adaptation of the traditional paper-and-pencil Trail Making Test within a three-dimensional, virtual environment. This innovative approach opens new avenues for enhancing neuropsychological and cognitive assessments. This study seeks to investigate the impact of different interaction modes—namely, eye tracking, head movement, and controller input—on performance accuracy and task completion time in the TMT-VR for a population of healthy young adults. Additionally, this study examines the influence of gaming abilities on these performance metrics, specifically completion time and accuracy. This study also aims to assess whether individuals with higher gaming proficiency demonstrate superior performance when utilizing head movement and eye tracking interaction modes compared to non-gamers and individuals with limited technological experience. Finally, given that the TMT-VR is a newly developed tool with no existing literature for direct comparison, this study will explore participants' perceived usability, acceptability, and overall user experience with the TMT-VR through two exploratory hypotheses.

Main hypotheses:

**H1.** *Eye tracking and/or head movement as interaction techniques will positively affect performance (accuracy and task completion time) on the TMT-VR part A and part B tasks.*

**H2.** *A higher level of gaming skill will positively affect performance (accuracy and completion time) on the TMT-VR part A and B tasks.*

**H3.** *The interaction between high gaming skill and eye tracking and/or head movement as interaction techniques will result in the best performance on the TMT-VR part A and part B tasks.*

Exploratory hypotheses:

**H4.** *The TMT-VR will demonstrate high usability, acceptability, and user experience, as defined by the developers of the respective questionnaires (e.g., SUTAQ, UEQ, and SUS).*

**H5.** *Positive associations with prior technology experiences and/or gaming skills will have significant positive effects on perceived usability, acceptability, and user experience of the TMT-VR.*

## 2. Materials and Methods

### 2.1. Participants

Participants were recruited using convenience and snowball sampling methods, primarily through the institution's Outlook mailing system, various social media platforms, and word of mouth. Informative posters with QR codes linking to the registration page were also distributed across the campus to facilitate sign-ups. Interested individuals voluntarily registered through an online calendar program. Eligible participants were young adults aged 18–35, all students at the American College of Greece, a private institution in Athens. To ensure the validity of this study, all participants were required to be proficient in English and free from neurodevelopmental disorders (e.g., ADHD and autism spectrum disorder), psychiatric disorders (e.g., depression and generalized anxiety disorder), learning difficulties (e.g., dyslexia and dyscalculia), neurocognitive impairments (e.g., mild cognitive impairment), neurological disorders (e.g., epilepsy and traumatic brain injury), neurodegenerative diseases (e.g., Alzheimers disease and Parkinsons disease), and physical disabilities (e.g., paraplegia and tetraplegia).

A total of 71 participants were included in this study, comprising 47.9% females ($n = 34$) and 52.1% males ($n = 37$). The participants' ages ranged from 18 to 35 years (M = 23.48, SD = 3.41), and their education levels varied between 12 and 23 years (M = 15.69, SD = 3.43). Regarding gaming ability, 47.9% of participants reported high gaming ability ($n = 34$), with 11.3% being female and 36.6% male. Conversely, 52.1% of participants reported low gaming ability ($n = 37$), with 36.6% being female and 15.5% male. All participants indicated a satisfactory level of proficiency in reading, writing, and understanding English.

## 2.2. VR Setup

In line with hardware suggestions to substantially mitigate the occurrence of cybersickness [32]. The VR setup utilized in this study was built around a high-performance PC paired with a Varjo Aero headset, Vive Lighthouse stations, SteamVR, noise-canceling headphones, and Vive controllers. The PC was equipped with an Intel64 Family 6 Model 167 Stepping 1 Genuine Intel processor running at approximately 3504 MHz, supported by 32 GB of RAM, ensuring smooth and responsive performance for VR applications.

The Varjo Aero headset was chosen for its exceptional visual clarity, featuring dual mini-LED displays with a resolution of 2880 × 2720 pixels per eye and a 115-degree horizontal field of view. The headset operates at a refresh rate of 90 Hz and includes custom-designed lenses to minimize distortion, providing an immersive and high-fidelity visual experience. Positional tracking was achieved using Vive Lighthouse stations, strategically positioned diagonally to cover a large tracking area and ensure precise tracking.

SteamVR was used as the primary platform for managing the VR experiences, offering robust room-scale tracking capabilities. To enhance auditory immersion, noise-cancelling headphones were employed, delivering high-fidelity audio with active noise cancellation to minimize external distractions. Interaction within the virtual environment was facilitated by Vive controllers, ergonomically designed with haptic feedback, grip buttons, trackpads or joysticks, and trigger buttons, ensuring a responsive and tactile user experience. The combination of these components provided a comprehensive and immersive VR setup, enabling detailed and precise Virtual Reality experiences for this study.

## 2.3. Development of the TMT-VR

The development of the TMT-VR adhered to the standards for Ergonomics of Human-System Interaction [58], which were essential in ensuring the system met the required ergonomic and usability criteria [58]. The design and development process followed the guidelines and recommendations for cognitive assessments in immersive VR, as outlined by [35], ensuring that the VR cognitive and psychomotor tasks developed were both scientifically valid and immersive. UCD principles played a pivotal role in the TMT-VR's development. UCD emphasizes creating interactive systems that prioritize the needs and capabilities of users, incorporating human factors and ergonomic considerations to maximize effectiveness, efficiency, and safety while minimizing performance issues [59]. In designing the TMT-VR tasks A and B, which require target selection from a distance, the method of ray casting was employed. Ray casting, a technique that simulates pointing with a laser pointer, was utilized across all interaction modes, including controller, headset, and eye-tracking, ensuring consistency and accuracy in user interactions.

Critical factors were considered to maintain the reliability of TMT-VR and its consistency with the traditional TMT. One such factor was the "Heisenberg effect", which posits that simultaneous confirmatory actions, such as pressing a button during pointing tasks, can impact performance accuracy [60]. To mitigate this, confirmatory button presses were eliminated in favor of automatic target selection after the participant focused on the target for 2 s. This approach aimed to reduce unintentional movements that could affect accuracy, as highlighted by [60]. Moreover, the development process addressed common VR interaction challenges, such as the Midas touch effect, where unintended targets might be selected during a search, and the occlusion effect, where targets might be obscured by other objects. These issues were mitigated by requiring a sustained focus of 2 s before selection and by designing the edges of occluded spheres to remain visible with slight head movements. These adjustments improved spatial awareness and task performance [61]. The selection of controllers over other modalities, such as joysticks or hand tracking, for ray casting in the TMT-VR was guided by UCD principles. Controllers were chosen due to their superior speed, precision, and user-perceived control, making them the most effective option for immersive VR environments [62]. To maintain the integrity of the experimental design, haptic feedback was excluded to avoid confounding variables, ensuring a consistent and reliable user experience. Additionally, the development of the TMT-VR took into consideration the accessibility needs of users with color vision deficiencies. The VR environment was designed using color-blind-friendly colors, as recommended by [63], to improve accessibility and ensure that all users, regardless of color vision status, could effectively engage with the tasks.

### 2.4. Materials

#### 2.4.1. Demographic Questionnaire

Participants completed a demographic questionnaire that collected information on age, gender, and proficiency in reading, writing, and understanding English. Additional questions confirmed the absence of neurodevelopmental disorders and learning difficulties diagnosed. Lastly, participants provided details on their technology use, particularly regarding the frequency and competency of their use of personal computers (PCs), smartphone applications, and VR tools.

#### 2.4.2. Gaming Skill Questionnaire (GSQ)

The Gaming Skill Questionnaire (GSQ) is a comprehensive tool designed to assess participants' gaming habits and skills across various video game genres. The questionnaire includes sections that evaluate gaming frequency and ability in specific genres such as Sports Games (SPGs), First-Person Shooting Games (FPSGs), Role-Playing Games (RPGs), Action-Adventure Games (AAGs), Strategy Games (STGs), and Puzzle Solving Games (PSGs). Participants rated their gameplay frequency and skill level on scales from 'Less than once a month' to 'Every day' for frequency, and from 'No experience' to 'Expert' for skill level. A Total Gaming Skill Score was derived by summing the scores across all genres, providing a comprehensive measure of participants' overall gaming involvement and expertise. The GSQ has demonstrated excellent psychometric properties, with high internal reliability (Cronbach's alpha ranging from 0.80 to 0.91) across all sections, indicating strong internal consistency. Furthermore, it exhibits robust construct validity, with strong convergent validity (item loadings ranging from 0.69 to 1) and excellent divergent validity, ensuring that the tool accurately assesses distinct dimensions of gaming skills within each genre [64,65].

#### 2.4.3. Cybersickness in VR Questionnaire (CSQ-VR)

The Cybersickness in VR Questionnaire (CSQ-VR) was used to assess the presence and severity of cybersickness symptoms in participants exposed to the VR environment. It evaluates nausea, vestibular issues, and oculomotor problems using a 7-point Likert scale. Participants responded to six questions, with two questions for each symptom category. Scores were summed to produce a total score, with a maximum possible score of 42 [8]. The psychometric properties of the CSQ-VR have shown good internal consistency, with reported Cronbach's alpha values ranging between 0.85 and 0.91, indicating high reliability. The CSQ-VR was validated against other cybersickness measurement tools and demonstrated strong construct validity, making it suitable for assessing cybersickness symptoms in Virtual Reality environments.

#### 2.4.4. The Trail Making Test in Virtual Reality (TMT-VR)

The Trail Making Test in Virtual Reality (TMT-VR) represents a significant advancement over the traditional paper-and-pencil version of the Trail Making Test. This VR adaptation was designed to harness the immersive capabilities of Virtual Reality, providing a more naturalistic and engaging environment for cognitive assessment. Participants were fully immersed in a 360-degree virtual environment, where numbered cubes were scattered within a field of view that extended across three dimensions, including the third axis (Z), offering a depth component that mirrors real-world spatial challenges (see Figure 1).

One of the key enhancements of the TMT-VR is its naturalistic approach to target selection. Unlike the traditional TMT, which relies on a two-dimensional paper format, the TMT-VR allows for the selection of targets through intuitive, real-world interactions such as eye-tracking, head movement, and the use of a VR controller. This approach closely mimics how individuals naturally interact with their environment, providing a more accurate assessment of cognitive functions like visual scanning, spatial awareness, and motor control.

The test administration in the TMT-VR is fully automated, including comprehensive tutorials that guide participants through the tasks. This automation ensures that all participants receive consistent instructions and guidance, eliminating the variability that can arise from human administrators. The scoring process is also automated and highly precise, removing the potential for human error in measuring performance. For example, the timing of task completion is automatically recorded, removing the need for manual start and stop actions typically required in traditional assessments, thereby providing a more accurate measure of cognitive processing speed.

A further enhancement is the fully randomized placement of targets within the virtual environment. Every time a participant undertakes the TMT-VR, the cubes appear in new, randomized positions, requiring a different spatial pattern of selection with each administration. This randomization ensures that the test has perfect test-retest reliability, as participants cannot simply memorize the location of targets from previous attempts. This feature is particularly important for longitudinal studies or repeated assessments, where the integrity of the test results must be maintained over multiple sessions.

Participants interacted with the virtual environment using ray-casting, where a virtual ray is projected from a specific point (e.g., the participant's eye, head, or controller) to select objects. The task required participants to alternate between numbered cubes in ascending order. Each cube had to be fixated upon for 1.5 s to confirm the selection, ensuring intentional and precise interactions. This two-phase process of aiming and confirming selection minimized errors due to inadvertent movements, which are common in traditional and virtual environments.

Visual and auditory feedback was provided to guide participants. Correct selections were highlighted in yellow, while incorrect ones were marked in red, accompanied by auditory cues. This immediate feedback allowed participants to quickly correct mistakes, reducing frustration and maintaining engagement. Previously selected cubes remained highlighted, helping participants track their progress and avoid reselecting cubes.

The TMT-VR included two tasks that corresponded to the traditional TMT-A and TMT-B (see Figure 1):

- TMT-VR Task A: Participants connected 25 numbered cubes in ascending numerical order (1, 2, 3,… 25). This task measured visual scanning, attention, and processing speed, mirroring the traditional TMT-A but within a more complex and immersive environment [39].
- TMT-VR Task B: Participants connected 25 cubes that alternated between numbers and letters in ascending order (1, A, 2, B, 3, C,… 13). This task assessed more complex cognitive functions, including task-switching ability and cognitive flexibility, reflecting the traditional TMT-B but within the enhanced VR setting [39].

Participants completed these tasks using three interaction modes as follows: eye-tracking, head movement, and VR controller. Each mode provided unique insights into the efficiency and accuracy of different input methods in VR. The accuracy metric, defined as the average distance from the center of the target cube during selection, provided a precise measure of how accurately participants were able to select the intended targets. Task completion time was automatically recorded, further ensuring precise and unbiased measurements of cognitive performance.

Overall, the TMT-VR offers a fully immersive, naturalistic, and randomized testing environment that not only replicates but also enhances the traditional TMT. Its automated administration, precise scoring, and naturalistic interaction modes make it a robust tool for cognitive assessment, offering greater ecological validity and reliability than traditional methods. This makes the TMT-VR particularly well-suited for diverse populations, including those with limited technological familiarity, while ensuring consistent and accurate cognitive evaluations across repeated assessments. A video presentation of each task of TMT-VR can be accessed using the following links: TMT-VR Task A (https://www.youtube.com/watch?v=npki7i4OnwY) and TMT-VR Task B (https://www.youtube.com/watch?v=immvIkOyVuA) (accessed on 1 September 2024).

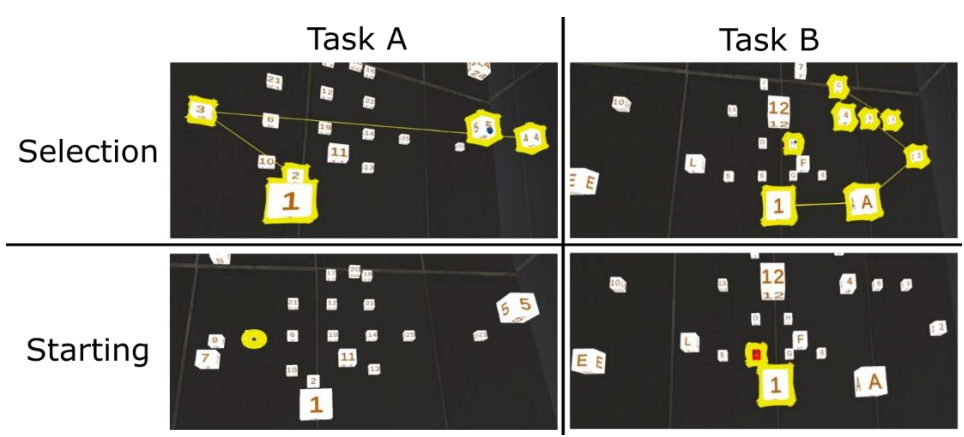

**Figure 1.** The Tasks of the TMT-VR. Task A (**left**), Task B (**right**), Starting Position (**bottom**), and Selection of Targets (**top**).

2.4.5. Acceptability Questionnaire (AQ)

The acceptability of the TMT-VR was assessed using an adapted version of the Service User Technology Acceptability Questionnaire (SUTAQ) [66]. This version consists of 10 items on a 6-point Likert scale, evaluating participants' satisfaction with the TMT-VR, including items such as "I am satisfied with the neuropsychological assessment of cognitive functions in Virtual Reality" [10]. The psychometric properties of the original SUTAQ have demonstrated good internal consistency, with Cronbach's alpha values ranging from 0.74 to 0.89 across various domains. These values indicate reliable measurement of user satisfaction and acceptability, making it suitable for assessing Virtual Reality systems in clinical and technological contexts.

### 2.4.6. User Experience Questionnaire—Short Version (UEQ-S)

The short version of the User Experience Questionnaire (UEQ-S) was utilized to measure the subjective user experience associated with the TMT-VR. This tool includes 26 items measured on a 7-point Likert scale with opposing terms (e.g., inefficient vs. efficient). The total score reflects participants' perceptions and feelings toward the TMT-VR [67]. The UEQ-S has demonstrated strong psychometric properties, with Cronbach's alpha values reported at 0.80 for pragmatic quality and 0.81 for hedonic quality, reflecting excellent internal consistency. Its construct validity was confirmed on multiple technology platforms, making it a reliable measure of user experience in Virtual Reality.

### 2.4.7. System Usability Scale (SUS)

The System Usability Scale (SUS) was employed to assess the usability of the TMT-VR. The SUS includes 10 items scored on a 5-point Likert scale, with scores converted into a total usability score out of 100. An example item is "I think I would like to use this tool frequently" [68]. The SUS is widely used for its simplicity and strong psychometric properties. It has demonstrated excellent internal consistency, with Cronbach's alpha values typically ranging from 0.85 to 0.92. In terms of validity, the SUS has shown to correlate well with other usability measures, and a SUS score above 68 is considered above average for usability, while scores above 80 are considered excellent.

### 2.5. Procedure

This study received ethical approval from the Ad-hoc Ethics Committee of the Psychology Department. Upon arrival at the psychology laboratory, participants were introduced to the VR equipment, including the headset, sensors, and controllers, and were briefed on the required bodily movements for the experiment's tasks. Following this introduction, participants reviewed and signed an informed consent form, and their eligibility based on the inclusion criteria was verified.

During the "pre-exposure" phase, participants completed demographic questions, a custom technology use questionnaire, the Gaming Skills Questionnaire (GSQ), and the Cybersickness in VR Questionnaire (CSQ-VR). Experimenters ensured participants' comfort and conducted eye tracking calibration using Varjo Base, where participants tracked a steady dot within the VR environment.

The experiment commenced with a specific executable file that initiated a mandatory second eye calibration phase involving tracking a moving dot. Participants completed the TMT-VR tasks (A and B) using eye-tracking, head movement, and controller interaction modes (see Figure 2). The sequence of interaction modes was counterbalanced to mitigate order effects and learning biases, with each mode preceded by a tutorial. Participants completed six tutorials, one for each task and interaction mode. Performance metrics, including accuracy and completion time, were recorded within a within-subjects experimental design.

Following the completion of the TMT-VR tasks, participants completed the Cybersickness in VR Questionnaire (CSQ-VR) again to assess post-exposure changes, along with the System Usability Scale (SUS), User Experience Questionnaire (UEQ-S), and Acceptance Questionnaire (AQ). Finally, participants were debriefed and thanked for their participation.

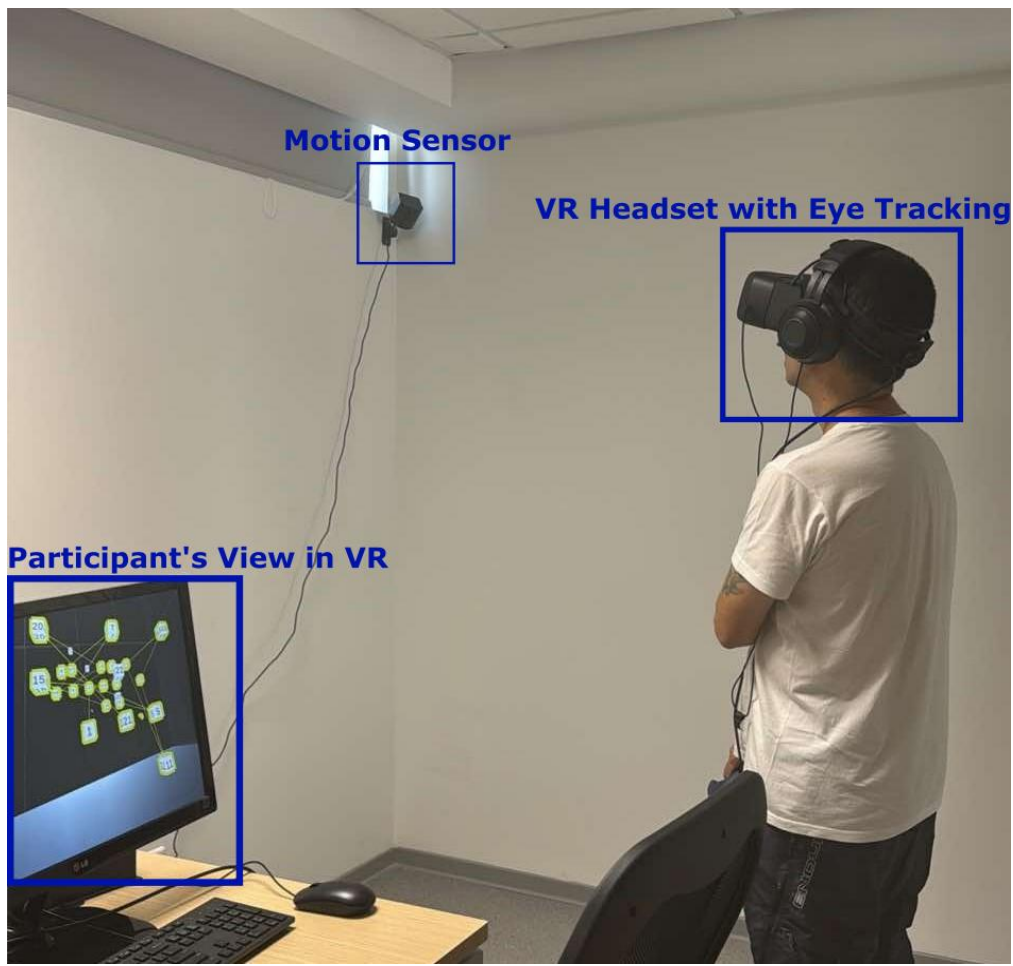

**Figure 2.** Participant performing TMT-VR using eye-tracking-based interaction.

*2.6. Statistical Analyses*

All statistical analyses were conducted using R software (v.4.4) [69] within the RStudio environment [70]. The dataset was first examined for normality using the Shapiro–Wilk test, with any non-normally distributed variables being transformed as necessary using the bestNormalize package (v.1.9.1) [71]. Descriptive statistics, including means, standard deviations (SDs), and ranges, were calculated for all demographic variables, task performance metrics, and user feedback scores. All data visualizations were created using the ggplot2 package (v.3.5.0) [72], which provided detailed graphical insights into the main effects and interactions from the ANOVA analyses. The plots were tailored to highlight key findings, including differences in task performance across interaction modes and gaming skill levels.

2.6.1. Correlation Analyses

Pearson's correlation coefficients were calculated to explore the relationships between demographic variables (e.g., age and education) and performance metrics (Task Time and Accuracy). Additionally, correlations were analyzed between user feedback scores (GSQ, UX, usability, and acceptability) and performance outcomes. The correlation matrices were visualized using the corrplot package (v.0.94) [73] to provide a clear and interpretable summary of these relationships.

2.6.2. Repeated Measures ANOVA

To assess the effects of interaction modes (eye tracking, headset, and controller) and gaming skill levels (high and low) on task time and accuracy for TMT-VR parts A and B, repeated measures ANOVAs were performed using the afex package (v.0.22) [74]. When the assumption of sphericity was violated, the Greenhouse–Geisser correction was applied. Post hoc pairwise comparisons were conducted using the Bonferroni correction to adjust for multiple comparisons. Effect sizes were reported as partial eta-squared ($\eta^2 p$), and a significance threshold was set at $p < 0.05$.

## 3. Results

### 3.1. Descriptive Statistics

The descriptive statistics are displayed in Table 1. All participants reported absent to very mild symptoms of cybersickness, with no significant differences between the pre- and post-exposure scores.

**Table 1.** Descriptive statistics.

| Variable | Mean (SD) | Range |
|---|---|---|
| Age | 23.48 (3.41) | 18–35 |
| Education | 15.69 (3.43) | 12–23 |
| PC | 9.97 (1.30) | 6–12 |
| SMART | 9.94 (1.59) | 4–12 |
| VR | 2.70 (1.30) | 2–8 |
| SPORT | 4.13 (2.23) | 2–12 |
| FPS | 4.73 (2.79) | 2–12 |
| RPG | 4.46 (3.04) | 2–12 |
| Action | 4.59 (2.82) | 2–12 |
| Strategy | 3.82 (2.75) | 2–12 |
| Puzzle | 4.06 (2.62) | 2–12 |
| GSQ | 25.79 (12.73) | 12–61 |
| UX | 46.76 (5.98) | 33–56 |
| Usability | 78.96 (9.58) | 60–98 |
| Acceptability | 49.34 (7.02) | 19–60 |

GSQ = Gaming Skills Questionnaire; PC = personal computer use; SMART = smartphone application use; VR = VR use; SPORT = Sports Games; FPS = First Person Shooting Games; RPG = Role Playing Games; Action = Action Adventure Games; Strategy = Strategy Games; Puzzle = Puzzle Games; GSQ = Gaming Skills Questionnaire; UX = User Experience Questionnaire; Usability = Usability Questionnaire; Acceptability = Acceptability Questionnaire.

### 3.2. Acceptability, Usability, and User Experience

The overall acceptability of the TMT-VR was high, with participants scoring a mean of 49.34 (SD = 7.02) out of a possible 60 on the Acceptability Questionnaire (see Table 1). This suggests that the majority of participants were satisfied with the neuropsychological assessment of cognitive functions in Virtual Reality. This high level of acceptability indicates that participants found the TMT-VR to be a satisfactory method for cognitive assessment and would be comfortable using similar tools in the future.

Usability was evaluated using the System Usability Scale (SUS), where participants provided a mean score of 39.48 (SD = 4.79) out of 50 (see Table 1). This indicates a generally positive perception of the system's usability. According to standard interpretations of SUS scores, this result suggests that the TMT-VR is considered to have good usability, allowing users to effectively interact with the virtual environment without significant difficulty. The User Experience Questionnaire-Short Version (UEQ-S) revealed that participants had a positive overall user experience, with a mean score of 46.76 (SD = 5.98) out of 56 (see Table 1). The majority of participants found the VR system to be efficient and enjoyable, suggesting that the TMT-VR provided a user-friendly and engaging platform for cognitive assessment.

The participants in this study provided generally favorable ratings for the acceptability, usability, and user experience of the VR setup used for the TMT-VR tasks. These findings suggest that the TMT-VR setup not only meets the technical requirements for usability but also aligns well with participant expectations in terms of comfort and effectiveness, thereby supporting its use in future cognitive assessment research.

### 3.3. Correlation Analyses

Correlational analyses revealed several significant associations among the performance metrics, demographics, and users' feedback (see Table 2). Age was negatively correlated with Task Time A and Task Time B and with Accuracy A and Accuracy B. Education revealed a negative association with Task Time B, Accuracy A and B, but a positive one with Task Time A. The GSQ negatively correlated with Task Time B and Accuracy A but revealed a positive relationship with Accuracy B. Additionally, User Experience (UX) was negatively correlated with Task Time A and both Accuracy

A and B. Lastly, Usability and Task Time A exhibited a positive association, while Acceptability demonstrated a positive relationship with Task Time A and Accuracy A but a negative one with Accuracy B.

**Table 2.** Correlations: performance with demographics and user evaluation.

|  | **Task Time A** | **Task Time B** | **Accuracy A** | **Accuracy B** |
|---|---|---|---|---|
| Age | −0.102 *** | −0.038 *** | −0.071 *** | −0.221 *** |
| Education | 0.002 ** | −0.038 *** | −0.097 *** | −0.038 *** |
| GSQ | 0.054 | −0.017 *** | 0.040 * | 0.004 ** |
| UX | −0.050 *** | 0.085 | −0.023 *** | 0.005 ** |
| Usability | 0.012 * | 0.118 | 0.112 | 0.102 |
| Acceptability | 0.036 * | 0.040 * | 0.008 ** | −0.093 *** |

$p < 0.05$ *, $p < 0.01$ **, $p < 0.001$ ***; GSQ = Gaming Skills Questionnaire; UX = User Experience Questionnaire; Usability = Usability Questionnaire; Acceptability = Acceptability Questionnaire.

In addition, age showed a positive correlation with User experience (UX) but negative correlations with Usability and Acceptability (see Table 3). Also, Education did not reveal any significance with User experience, yet negative correlations were detected with Usability and Acceptability. Also, GSQ did not reveal a significant correlation with User experience and Usability, but a negative correlation with Acceptability was revealed. Lastly, PC and smart phone use were positively correlated with User Experience, Usability, and Acceptability (see Table 3). Therefore, the second exploratory hypothesis is partially confirmed.

**Table 3.** Correlation analysis of demographic variables, performance metrics, and user feedback scores in the TMT-VR study.

|  | **UX** | **Usability** | **Acceptability** |
|---|---|---|---|
| Age | 0.031 * | −0.148 *** | −0.026 *** |
| Education | 0.081 | −0.003 *** | −0.012 *** |
| GSQ | 0.193 | 0.226 | −0.014 *** |
| PC | 0.230 | 0.253 | 0.040 * |
| SMART | 0.187 | 0.292 | 0.135 |

$p < 0.05$ *, $p < 0.01$ **, $p < 0.001$ ***; GSQ = Gaming Skills Questionnaire; PC = personal computer use; SMART = smartphone application use.

*3.4. Mixed Measures ANOVA Analyses*

The results of the analyses for Accuracy and Task Time across different gaming levels and interaction modes are presented in Table 4. This table summarizes the mean performance metrics, including the standard deviation and range, for each condition. The data provide a comprehensive overview of how different interaction modes (eye tracking, headset, and controller) influenced the participants' accuracy and task completion times for both Task A and Task B. The table also highlights the impact of gaming skill levels (high and low) on these performance metrics.

**Table 4.** Accuracy and task time by gaming level and interaction mode.

| Task | Metric | Gaming Level | Interaction Mode | Mean (SD) | Range (Min–Max) |
|------|--------|--------------|------------------|-----------|-----------------|
| A | Accuracy | High | Eye | 15.6 (0.249) | 15.2–16.2 |
| | | | Hand | 15.7 (0.180) | 15.3–16.0 |
| | | | Head | 15.6 (0.190) | 15.2–16.0 |
| | | Low | Eye | 15.5 (0.233) | 15.0–15.9 |
| | | | Hand | 15.8 (0.256) | 15.3–16.3 |
| | | | Head | 15.6 (0.215) | 15.1–15.9 |
| | Accuracy | Overall | Eye | 15.5 (0.255) | 15.0–16.2 |
| | | | Hand | 15.7 (0.222) | 15.3–16.3 |
| | | | Head | 15.6 (0.202) | 15.1–16.0 |
| | Accuracy | High | Overall | 15.6 (0.214) | 15.2–16.2 |
| | | Low | Overall | 15.6 (0.265) | 15.0–16.3 |
| B | Accuracy | High | Eye | 15.6 (0.270) | 14.9–16.2 |
| | | | Hand | 15.6 (0.238) | 15.2–16.2 |
| | | | Head | 15.6 (0.226) | 15.1–16.2 |
| | | Low | Eye | 15.5 (0.189) | 15.2–15.9 |
| | | | Hand | 15.7 (0.232) | 15.3–16.1 |
| | | | Head | 15.5 (0.254) | 15.1–16.1 |
| | Accuracy | Overall | Eye | 15.5 (0.232) | 14.9–16.2 |
| | | | Hand | 15.7 (0.235) | 15.2–16.2 |
| | | | Head | 15.6 (0.241) | 15.1–16.2 |
| | Accuracy | High | Overall | 15.6 (0.244) | 14.9–16.2 |
| | | Low | Overall | 15.6 (0.236) | 15.1–16.1 |
| A | Task Time | High | Eye | 73.9 (11.487) | 51.8–103.1 |
| | | | Hand | 80.8 (15.681) | 55.6–116.5 |
| | | | Head | 67.2 (15.946) | 42.3–118.2 |
| | | Low | Eye | 74.8 (17.421) | 42.3–113.8 |
| | | | Hand | 82.3 (16.926) | 53.8–120.0 |
| | | | Head | 72.2 (16.232) | 45.8–119.5 |
| | Task Time | Overall | Eye | 74.4 (14.782) | 42.3–113.8 |
| | | | Hand | 81.6 (16.242) | 53.8–120.0 |
| | | | Head | 69.8 (16.172) | 42.3–119.5 |
| | Task Time | High | Overall | 74.0 (15.418) | 42.3–118.2 |
| | | Low | Overall | 76.4 (17.256) | 42.3–120.0 |
| B | Task Time | High | Eye | 84.1 (17.085) | 46.1–108.8 |
| | | | Hand | 90.5 (17.244) | 62.9–125.8 |
| | | | Head | 82.0 (20.783) | 45.6–118.5 |
| | | Low | Eye | 85.0 (20.408) | 49.5–130.9 |
| | | | Hand | 95.2 (23.996) | 48.7–150.9 |
| | | | Head | 82.6 (21.638) | 48.5–126.2 |
| | Task Time | Overall | Eye | 84.6 (18.762) | 46.1–130.9 |
| | | | Hand | 92.9 (21.018) | 48.7–150.9 |

| | | Head | 82.3 (21.083) | 45.6–126.2 |
|---|---|---|---|---|
| Task Time | High | Overall | 85.6 (18.625) | 45.6–125.8 |
| | Low | Overall | 87.6 (22.538) | 48.5–150.9 |

Accuracy = average distance from the center in centimeters; Task Time = time required to complete the task in seconds.

### 3.4.1. Task Time

A repeated measures ANOVA was conducted to assess the influence of interaction modes (eye tracking, headset, and controller) on task time for TMT-VR parts A and B. Results revealed a significant main effect of interaction modes, $F(2, 207) = 10.63$, $p < 0.001$, $\eta^2 p = 0.09$ for Task Time A, and $F(2, 207) = 5.63$, $p = 0.004$, $\eta^2 p = 0.05$ for Task Time B. Post hoc pairwise comparisons revealed significant differences in Task Time A between the eye and controller interaction modes ($p = 0.002$) and between the head and the controller ($p < 0.001$; see Figure 3). Regarding Task Time B, Post hoc pairwise comparisons revealed similar data to Task Time A, with main effects presented between the eye and controller interaction modes ($p = 0.002$) and between the head and the controller ($p < 0.001$; see Figure 4). Notably, the headset mode exhibited superior performance in task completion time compared to the controller mode. However, there was no statistically significant difference in task completion time between the headset and eye tracking modes. Therefore, the first hypothesis is confirmed. These findings suggest that participants' performance in task completion time was not significantly influenced by their gaming skill level. Similarly, a repeated measures ANOVA analysis revealed that there were no significant interactions between gaming skill and interaction mode (eye, headset, and controller) both for Task Time A, that is $F(2, 207) = 0.668$, $p = 0.514$, $\eta^2 p = 0.001$, and for Task Time B, that is, $F(2, 207) = 0.143$, $p = 0.866$, $\eta^2 p = 0.001$. Therefore, the second and third hypotheses are rejected.

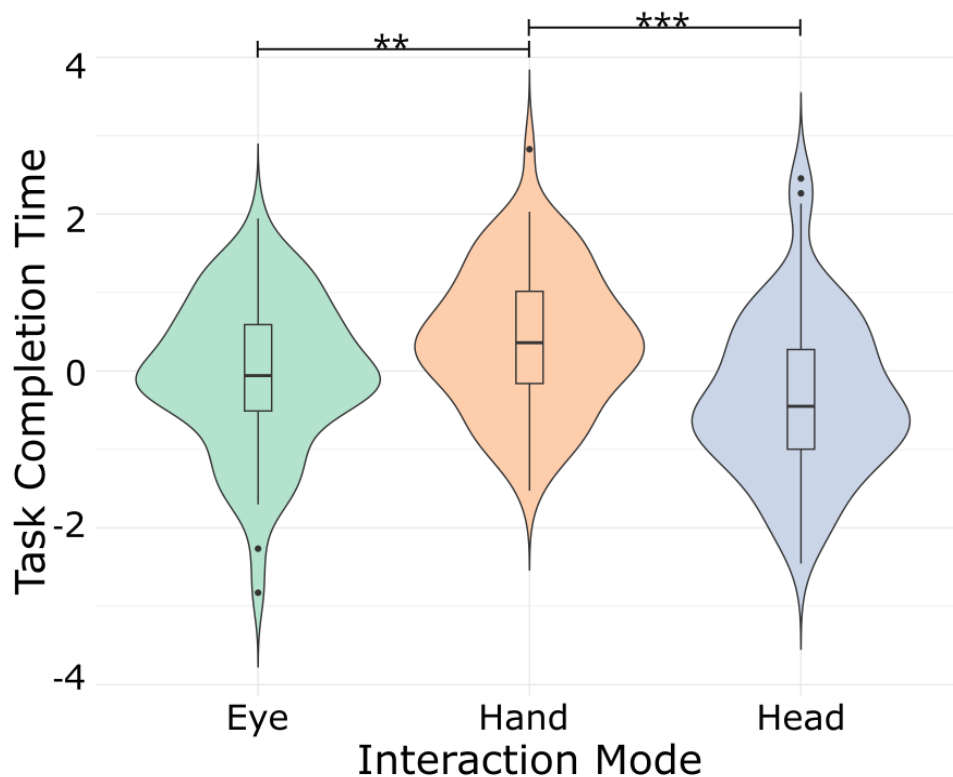

**Figure 3.** Task completion time of TMT-VR Task A per interaction mode and gaming skill. Task completion time is displayed as a Z score.

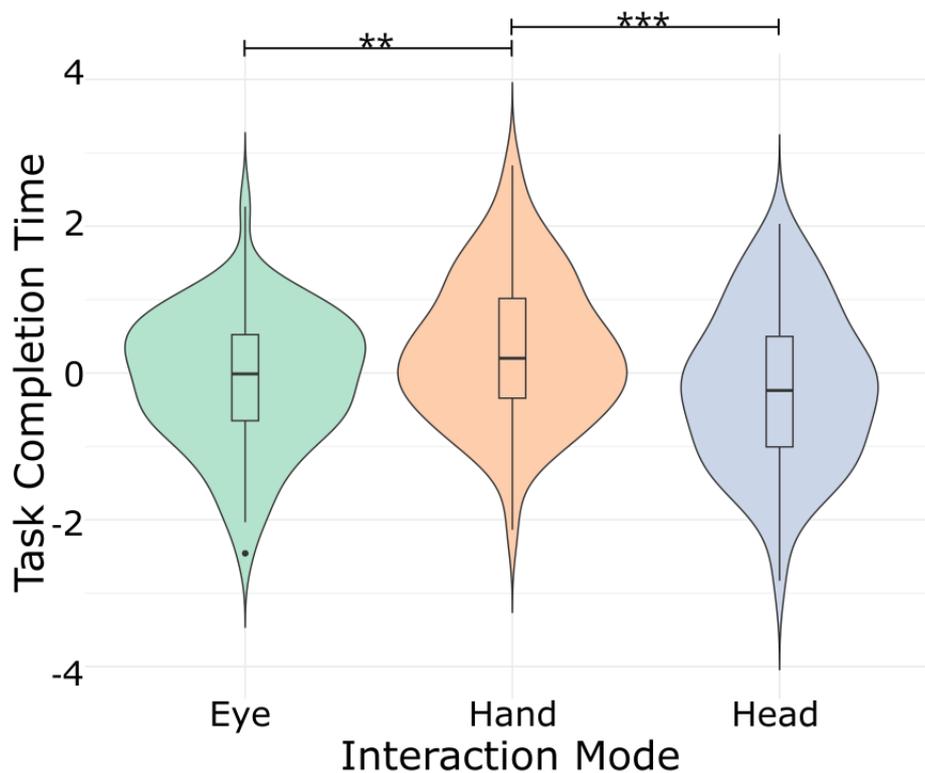

**Figure 4.** Task completion time of TMT-VR Task B per interaction mode and gaming skill. Task completion time is displayed as a Z score.

### 3.4.2. Task Accuracy

To examine the impact of different interaction modes (eye tracking, headset, and controller) and gaming skills on TMT-VR parts A and B Task Accuracy, repeated measures ANOVA were conducted. Results revealed a statistically significant interaction for TMT-VR Accuracy A between gaming skills and interaction modes as follows: $F_{(2, 207)}$ = 4.47, $p = 0.012$, $\eta^2 p = 0.04$ (see Figure 5). Additionally, the interaction mode also presented significant results as follows: $F_{(2, 207)}$ = 16.59, $p < 0.001$, $\eta^2 p = 0.14$ (see Figure 6). Participants in the high gaming skills group provided significant results in relation to the interaction modes as follows: $F_{(1.83, 60.24)}$ = 4.53, $p = 0.02$, $\eta^2 p = 0.06$. Post hoc pairwise comparisons revealed a significant difference between the head and the controller ($p < 0.001$). Similarly, the low gaming skills group provided significant results in relation to interaction modes as follows: $F_{(1.98, 71.42)}$ = 15.25, $p < 0.001$, $\eta^2 p$ = 0.21, and Post hoc pairwise comparisons showed significant differences between eye tracking and the controller ($p < 0.001$) and between the head and the controller ($p < 0.001$).

The statistical analyses regarding Accuracy B revealed a statistical significance of interaction mode as follows: $F_{(2, 207)}$ = 4.18, $p = 0.017$, $\eta^2 p = 0.04$. Regarding the Accuracy TMT-VR B, the effects of gaming skill (high/low) were not statistically significant as follows: $F_{(2, 207)}$ = 1.09, $p = 0.337$, $\eta^2 p = 0.001$. However, Post hoc pairwise comparisons revealed a significant result between eye tracking and the controller ($p = 0.02$; see Figure 7). Notably, the eye tracking mode demonstrated superior accuracy compared to the controller mode. However, there was no statistically significant difference in task accuracy between the eye and headset modes ($p = 0.103$), suggesting comparable performance between these two modalities. Overall, these findings highlight that the eye tracking mode emerged as the most accurate, outperforming the controller mode while exhibiting no significant differences compared to the headset mode.

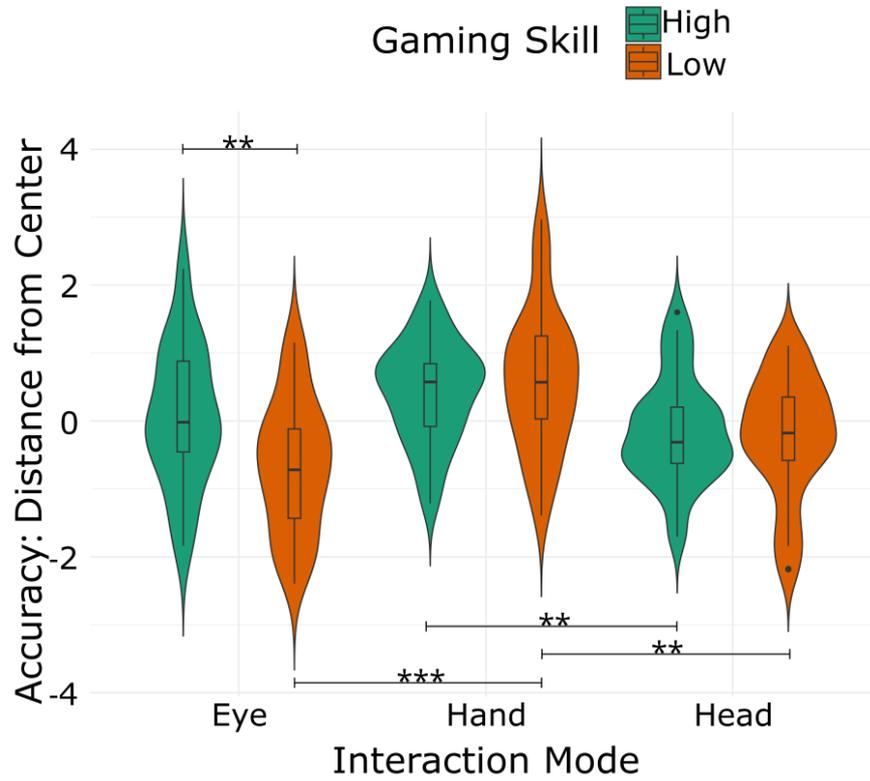

**Figure 5.** Accuracy in TMT-VR Task A per interaction mode and gaming skill. Accuracy is displayed as a Z score of the distance from the center.

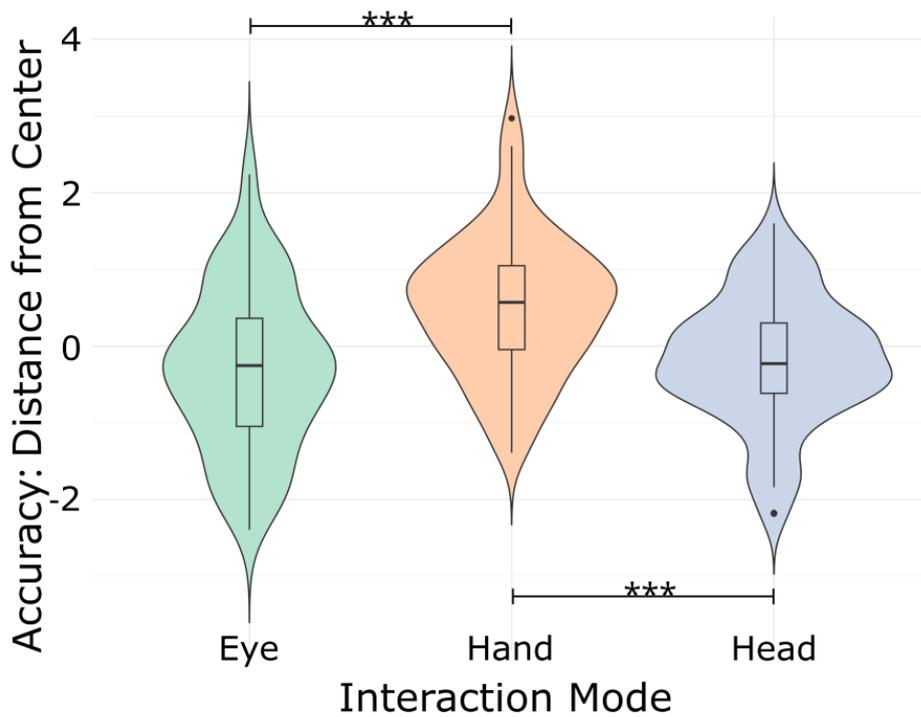

**Figure 6.** Accuracy in TMT-VR Task A per interaction mode. Accuracy is displayed as a Z score of the distance from the center.

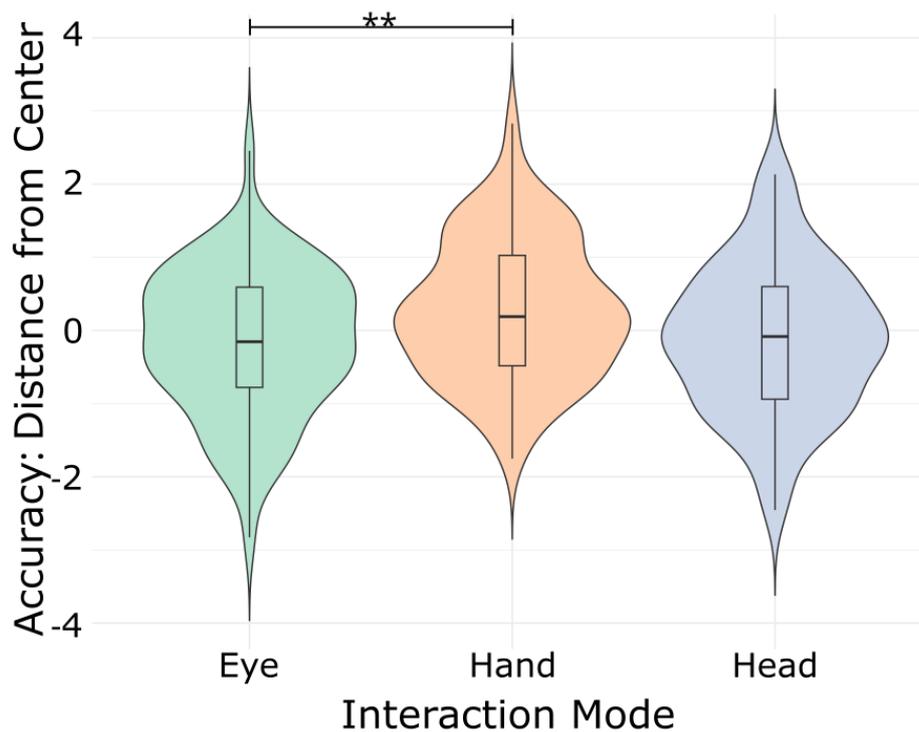

**Figure 7.** Accuracy in TMT-VR Task B per interaction mode. Accuracy is displayed as a Z score of the distance from the center.

## 4. Discussion

The current study aimed to develop and evaluate a Virtual Reality (VR) adaptation of the traditional Trail Making Test (TMT), focusing on task completion time and accuracy among healthy young adults. This VR adaptation, TMT-VR, was designed to accommodate various interaction modes—eye-tracking, head movement, and controller interaction—while also examining the influence of prior gaming and technology skills on performance. Additionally, this study assessed the usability, user experience, and acceptability of the TMT-VR as a neuropsychological assessment tool. The findings of this study contribute to the growing body of literature on the integration of VR into cognitive assessments, offering insights into how different interaction modes and user characteristics influence performance and user experience.

### 4.1. Usability and User Experience

The TMT-VR received high ratings for usability, user experience, and acceptability, reflecting the successful application of UCD principles in its development. UCD emphasizes creating systems that are not only functional but also tailored to meet the specific needs and preferences of users, thereby enhancing overall satisfaction and effectiveness. The positive evaluations observed across all participants, regardless of their gaming or technology experience, indicate that the TMT-VR is accessible and user-friendly. This is a significant achievement, as it suggests that the TMT-VR can be effectively used in diverse populations, including those who may not be familiar with VR technology [2]. Furthermore, those findings indicate that TMT-VR has the potential to increase accessibility in cognitive evaluations by adhering to the distinct requirements and preferences of diverse populations. This is crucial for groups that may otherwise face barriers in conventional cognitive assessment techniques, such as individuals with physical impairments and older adults. This broad applicability is crucial for neuropsychological assessments, which often need to cater to a wide range of users, including clinical populations and older adults who may have limited technological exposure [75].

The development of the TMT-VR adhered closely to the ISO 9241-210:2019 standards for Ergonomics of Human-System Interaction, which provide comprehensive guidelines for designing interactive systems that are efficient, effective, and satisfying for users [58]. By aligning the design of the TMT-VR with these standards, the developers ensured that the tool not only met the technical requirements for usability but also addressed the ergonomic needs of users, reducing the likelihood of discomfort and cognitive overload during the assessment process. This alignment with ISO standards is critical in the context of VR, where user comfort and interaction quality are paramount to prevent issues like cognitive fatigue and physical discomfort, which can negatively impact the user experience and the validity of the assessment outcomes [76].

Incorporating specific features to mitigate common VR issues, such as the Midas touch effect and cybersickness, further contributed to the positive user experience reported in this study. The Midas touch effect, which occurs when a system interprets unintended user actions as intentional, can be particularly problematic in VR environments where precise control is necessary [61]. To address this, the TMT-VR implemented a two-second confirmation delay for target selection, allowing users to correct potential errors before finalizing their actions. This feature is consistent with best practices in VR interface design, which recommend implementing safeguards against inadvertent actions to enhance user control and reduce frustration [60].

Cybersickness, another common issue in VR, can manifest as symptoms such as dizziness, nausea, and disorientation, often resulting from discrepancies between visual and vestibular inputs [77]. The TMT-VR's design includes measures to minimize these effects, such as optimizing the frame rate and implementing smooth, consistent motion within the virtual environment. These strategies are in line with the recommendations for reducing VR-induced symptoms, which emphasize the importance of maintaining a high level of visual fidelity and ensuring that user movements are accurately reflected in the virtual space [78]. The successful mitigation of cybersickness in the TMT-VR is particularly important for its application in clinical settings, where users may already be vulnerable to discomfort or disorientation [2].

Furthermore, the TMT-VR's design carefully considered the user experience by incorporating elements that enhance immersion and reduce potential barriers to effective interaction. For instance, the system's use of intuitive interaction modes, such as eye tracking and head movement, aligns with natural human behaviors and reduces the physical and cognitive demands placed on users [45]. These interaction modes are particularly beneficial for older adults or individuals with motor impairments, as they require less physical effort and are easier to master than traditional controller-based interactions [79]. The integration of these features into the TMT-VR not only improves user satisfaction but also enhances the tool's ecological validity, ensuring that the virtual tasks more closely replicate real-world cognitive challenges [35].

In summary, the high usability, user experience, and acceptability ratings of the TMT-VR reflect the successful implementation of UCD principles and ISO standards in its design. By addressing both the technical and ergonomic needs of users, the TMT-VR has demonstrated its potential to be an effective and user-friendly tool for cognitive assessment in diverse populations. The careful attention to mitigating common VR issues, along with the incorporation of naturalistic interaction modes, has resulted in a system that is both accessible and satisfying to use. These findings contribute to the growing body of evidence supporting the use of VR technology in neuropsychology and underscore the importance of user-centered design in the development of effective and inclusive assessment tools [35,68].

## 4.2. Enhancing Performance

### 4.2.1. Task Completion Time

The analysis of task completion time for TMT-VR Parts A and B revealed significant differences among the interaction modes. Specifically, the headset mode consistently outperformed the controller mode, with participants completing tasks more quickly. This finding aligns with previous research that suggests head movement-based interaction offers a more intuitive and less cognitively demanding method for interacting with virtual environments, as it mimics natural head movements used in real-world tasks [79]. Moreover, the lack of significant differences between the headset and eye tracking modes suggests that both are efficient for task completion, with the headset being slightly quicker, likely due to its more stable and precise control in VR environments [80]. The absence of significant interactions between gaming skill and interaction mode further indicates that the advantages of the headset and eye tracking modes are consistent across different levels of gaming experience, reinforcing the versatility of these interaction methods in VR cognitive assessments.

### 4.2.2. Task Accuracy

In terms of task accuracy, this study found that high-gamers performed more accurately using the headset in TMT-VR Part A, while low-gamers were more accurate with the eye tracking modality. This finding suggests that individuals with more gaming experience may be better at fine-tuned interactions, such as those required by head movements, while those with less experience might benefit more from the naturalistic and direct control offered by eye tracking [45]. For TMT-VR Part B, eye tracking was generally more accurate than the controller, with no significant differences between the eye tracking and headset modes. These findings align with the existing literature that highlights the precision of eye tracking in tasks requiring accurate target selection, particularly in scenarios involving depth and spatial awareness [79]. The effectiveness of eye tracking and head movement as interaction modes in VR-based cognitive assessments is further supported by their ability to provide natural and intuitive control, minimizing the cognitive load on participants [2].

## 4.3. Gaming Skills

In contrast to previous studies, gaming skills did not significantly enhance task performance in the TMT-VR, challenging the assumption that gamers would inherently perform better in VR-based cognitive tasks. Although high-skilled gamers showed slight advantages in specific conditions—such as improved accuracy in certain interaction modes—these benefits were inconsistent across tasks and interaction modes. This suggests that the cognitive advantages conferred by gaming may be more context-dependent than previously thought, particularly in novel or complex VR environments. One possible explanation is that the benefits of gaming experience are task-specific; gamers may excel in scenarios that closely align with video game challenges, such as rapid target tracking or spatial navigation. When the tasks in the TMT-VR do not align with these skills, the manifestation of gaming-related benefits may be limited. Additionally, the absence of a performance boost could be linked to cognitive load and adaptability, as skilled gamers typically manage higher cognitive demands and adapt quickly to new challenges. However, if the VR tasks do not sufficiently challenge their abilities, the expected superiority may not emerge. Finally, individual differences among gamers, including cognitive style, learning speed, and previous VR exposure, may contribute to variability in performance outcomes, further complicating the relationship between gaming experience and VR task performance [26,81].

This study's findings contrast with earlier research that suggests gamers typically demonstrate superior performance in executive functions, such as working memory and task switching. In these studies, gamers often outperformed non-gamers in tasks requiring quick adaptation to changing rules or environments, likely due to their extensive experience with the fast-paced, dynamic nature of many video games. However, the findings of this study align with recent evidence indicating that the cognitive benefits of gaming experience may not extend uniformly to all types of cognitive tasks, especially those that are more complex, novel, or outside the typical scope of gaming activities [2].This suggests that the cognitive skills developed through gaming are specialized and may not generalize to other areas as effectively as previously assumed.

In contrast to traditional gaming environments, Virtual Reality (VR) places unique sensory and perceptual demands on users. Operating within a fully immersive 3D space, users must engage distinct perceptual and motor skills, such as depth perception, spatial orientation, and proprioception. These skills are critical for navigating and interacting with VR environments and may not be as developed in gamers accustomed to conventional 2D interfaces.

Consequently, the advantages typically held by experienced gamers in traditional settings may be diminished when transitioning to VR, as the required cognitive and motor proficiencies differ significantly [82].

Overall, the results of this study suggest that performance in the TMT-VR is largely independent of prior gaming experience, making the tool accessible and effective for a broader range of users, including those with limited gaming or technological backgrounds. This finding is particularly encouraging as it implies that the TMT-VR can be used in diverse populations, such as older adults or individuals less familiar with technology, without compromising the validity of the assessment. The tool's usability across different demographic groups enhances its ecological validity, positioning the TMT-VR as a valuable instrument in neuropsychological assessments where the diverse cognitive abilities and backgrounds of users must be considered. These results underscore the importance of designing VR-based cognitive tools that are inclusive and not overly reliant on specific user experiences, ensuring broad applicability and effectiveness across various populations.

### 4.4. Interaction Modes

The comparison of interaction modes demonstrated that both eye tracking and headset interactions offer distinct advantages over traditional controller-based interaction. Eye tracking proved particularly effective for tasks requiring depth perception, where precise visual focus on distant targets is essential [79]. The headset, however, provided a balance of speed and precision, making it an effective mode for tasks that require rapid and accurate responses [80]. These findings suggest that the choice of interaction mode can be tailored to the specific needs of the user and the nature of the task, with both eye tracking and headset interactions providing naturalistic and intuitive options for VR assessments [2]. This adaptability highlights the potential of these interaction modes to enhance the accessibility and effectiveness of VR-based cognitive assessments.

The finding that both eye tracking and head movement modes outperformed controller-based interaction in terms of accuracy and task completion time presents important implications for the future of VR interaction design. While controllers are currently the dominant mode of interaction in most VR applications, this study suggests that they may not always be the most efficient or accurate method for specific tasks, particularly those requiring precise targeting and rapid responses. One possible explanation for the superior performance of eye tracking and head movement modes lies in their alignment with natural human behaviors. Eye tracking and head movement are intrinsic to how we interact with the real world, where we use our eyes and heads to focus on and select objects. This natural alignment likely reduces cognitive load, allowing users to interact with the virtual environment more intuitively and efficiently than with controllers, which require hand-eye coordination and manual dexterity. Additionally, these modes may offer ergonomic benefits by reducing physical demand compared to controllers, which could help maintain accuracy and speed over longer periods, particularly in tasks that require sustained precision.

These findings have significant implications for the design and development of future VR software. While controllers have been the focus due to their widespread use and adaptability, the potential of eye tracking and head movement to provide more efficient and user-friendly interactions deserves greater attention. As VR technology continues to evolve, developers might consider integrating or even prioritizing these alternative interaction modes, offering users more intuitive and less physically demanding ways to interact with virtual environments. Future research could explore the development of hybrid interaction systems that combine the strengths of all these modes, allowing users to choose the most appropriate method for their specific tasks. In conclusion, while controller-based interaction remains prevalent in VR applications, this study demonstrates that alternative modes such as eye tracking and head movement may offer significant benefits in terms of accuracy and task completion time, suggesting a potential shift in how we think about VR interaction design.

### 4.5. Limitations and Future Directions

While the present study offers valuable insights into the effectiveness of the TMT-VR as a neuropsychological assessment tool, a few limitations should be considered, which also pave the way for future research. Although the sample size was adequate for the statistical analyses conducted, this study primarily involved healthy young adults with relatively high levels of education. This demographic homogeneity, while useful for establishing baseline performance, limits the generalizability of the findings to other populations, such as older adults or individuals with cognitive impairments. Future studies should aim to include a more diverse sample, encompassing a broader age and education range, as well as clinical populations, to better understand the TMT-VR's applicability and reliability across various groups. Once validated with older adults, TMT-VR could provide accessible cognitive assessments tailored to this population's specific needs. Furthermore, although the primary focus of this initial study was on user experience and interaction modes, these findings lay the groundwork for subsequent studies that can evaluate the cognitive validity of the TMT-VR more directly. Lastly, future research could incorporate feature importance analysis to identify which attributes have the greatest impact on performance across interaction modes.

## 5. Conclusions

This study evaluated the TMT-VR, demonstrating its effectiveness as a cognitive assessment tool that leverages immersive VR technology. The findings reveal that both eye tracking and head movement interaction modes significantly outperformed the traditional controller-based interaction in terms of accuracy and task completion time. This suggests that more naturalistic and intuitive interaction methods, such as eye tracking and head movement, may offer superior performance in VR-based cognitive assessments, potentially shifting the focus away from controllers, which are currently predominant in VR applications.

Importantly, this study also found that prior gaming experience did not significantly enhance performance on the TMT-VR, indicating that the tool is accessible and effective for users regardless of their familiarity with gaming or technology. High ratings of usability, acceptability, and user experience further validate the TMT-VR as a reliable and user-friendly assessment tool suitable for diverse populations.

These results underscore the potential of VR to revolutionize cognitive assessments by providing engaging, effective, and user-friendly platforms that replicate traditional tests while offering new possibilities for neuropsychological evaluation. As VR technology continues to advance, further refinement of tools like the TMT-VR, with a focus on inclusivity and ergonomic optimization, will be essential to maximize their impact in both clinical and research settings.

**Author Contributions:** Conceptualization, E.G., P.V. and P.K.; methodology, E.G., P.V. and P.K.; software, P.K.; validation, E.G., P.V., R.K., I.K., C.N. and P.K.; formal analysis, E.G., P.V. and P.K.; investigation, E.G., P.V. and R.K.; resources, I.K., C.N. and P.K.; data curation, E.G., P.V. and R.K.; writing—original draft preparation, E.G., P.V., R.K. and P.K.; writing—review and editing, E.G., P.V., R.K., I.K., C.N. and P.K.; visualization, P.K.; supervision, P.K.; project administration, I.K., C.N. and P.K.; funding acquisition, I.K., C.N. and P.K. All authors have read and agreed to the published version of the manuscript.

**Funding:** This research received no external funding.

**Institutional Review Board Statement:** This study was conducted in accordance with the Declaration of Helsinki and approved by the Ad-hoc Ethics Committee of the Psychology Department of the American College of Greece (KG/0224, 28 February 2024).

**Informed Consent Statement:** Informed consent was obtained from all subjects involved in this study.

**Data Availability Statement:** The data presented in this study are available on request from the corresponding author. The data are not publicly available due to ethical approval requirements.

**Acknowledgments:** This study received financial support by the ACG 150 Annual Fund. Panagiotis Kourtesis designed and developed the TMT-VR.

**Conflicts of Interest:** The authors declare no conflicts of interest.